%% file: p16_02_11.tex
\documentclass[aps,pra,superscriptaddress,preprint]{revtex4-1}

\usepackage{amsmath,amssymb}
\usepackage{graphicx}% Include figure files
\usepackage{caption}
\usepackage{color}% Include colors for document elements
\usepackage{dcolumn}% Align table columns on decimal point
\usepackage{bm}% bold math
\usepackage{graphicx}
\usepackage{color}
\usepackage{ulem}
\usepackage{mathtools}

\input epsf

\input dft

\input rotate

\def\bei{\begin{itemize}}
\def\eei{\end{itemize}}
\def\2c{_{2\sss C}}

\def\FG{^{{\sss FG}}}
\def\inter{_{{\sss INT}}}
\def\GS{_{{\sss GS}}}

\begin{document}

\title{Density-Matrix Propagation Driven by Semiclassical Correlation}

\author{Peter Elliott}
\affiliation{Max-Planck-Institut f\"ur Mikrostrukturphysik, Weinberg 2, D-06120 Halle, Germany}
\author{Neepa T. Maitra}
\affiliation{Department of Physics and Astronomy, Hunter College and the City University of New York, 695 Park Avenue, New York, New York 10065, and Physics Program, The Graduate Center, CUNY, New York, New York 10016, USA}

%%%%%%%%%%%%%%%%%%%%%%%%%%%%%%%%%%%%%%%%%%%%%%%%%%%%

\begin{abstract}
Methods based on propagation of the one-body reduced density-matrix
hold much promise for the simulation of correlated many-electron
dynamics far from equilibrium, but difficulties with finding good
approximations for the interaction term in its equation of motion have
so far impeded their application. These difficulties include the
violation of fundamental physical principles such as energy
conservation, positivity conditions on the density, or unchanging
natural orbital occupation numbers. We review some of the recent efforts
to confront these problems, and explore a semiclassical
approximation for electron correlation coupled to time-dependent
Hartree-Fock propagation. We find that this approach captures changing
occupation numbers, and excitations to doubly-excited states,
improving over TDHF and adiabatic approximations in density-matrix
propagation. However, it does not guarantee $N$-representability of
the density-matrix, consequently resulting sometimes in violation of
positivity conditions, even though a purely semiclassical treatment
preserves these conditions.
\end{abstract}

\maketitle

%%%%%%%%%%%%%%%%%%%%%%%%%%%%%%%%%%%%%%%%%%%%%%%%%%%%

\section*{INTRODUCTION} 

\label{sec:intro}
Time-resolved dynamics of electrons in atoms, molecules, and solids
are increasingly relevant for a large class of problems today. The
electrons and ions are excited far from their ground states in
photo-induced processes such as in photovoltaics or laser-driven
dynamics, and control of the dynamics on the attosecond time-scale is now
experimentally possible.  To theoretically model these
processes an adequate accounting of electron correlation is
required. Clearly solving the full time-dependent Schr\"odinger equation (TDSE) is impossible for more than a
few electrons, and, moreover, the many-electron wavefunction contains
much more information than is needed. Most observables of interest
involve one- or two-body operators, suggesting that a description in
terms of reduced variables would be opportune: in particular,
obtaining directly just the one- and two-body time-dependent reduced density matrices
(TD RDMs)~\cite{RDMbook} would enable us to obtain any one- or two-body observable
(e.g. electron densities, momentum profiles, double-ionization
probabilities, etc).  Even simpler, the theorems underlying
time-dependent density functional theory (TDDFT), prove that {\it any}
observable can be obtained from knowledge of simply the one-body
density. However, hiding in any of these reduced descriptions is the
complexity of the full many-body interacting electron problem, in the
form of reconstruction functionals for the RDM case and
exchange-correlation potentials as well as observable-functionals in
the TDDFT case. In practice, these terms must be approximated, and
intense research has been underway in recent years to determine
approximations that are accurate but practically
efficient. The temptation to simply use approximations that were
developed for the ground-state cases, whose properties for the
ground-state have been well-studied and understood, has proved
profitable in some cases giving, for example, usefully accurate
predictions of excitation spectra~\cite{EFB07,TDDFTbook2,PG15}.  But when used for
non-perturbative dynamics, these same approximations can become rapidly
unreliable. These problems will be reviewed in the next section.

Solving the full TDSE scales exponentially with the number of
electrons in the system, while the computational cost of propagating
RDMs is in principle independent of the system-size. Classical
dynamics of many-body systems on the other hand scales linearly with
the number of particles, which raises the question of using
semiclassical approaches to many-electron systems. Usually used for
nuclear dynamics, a semiclassical wavefunction is built using
classical dynamical information alone, in particular the phase arises
from the classical action along the
trajectory~\cite{V28,S81,H81,M98,TW04,K05}; in this way semiclassical
methods can capture essential quantum phenomena such as interference,
zero-point energy effects, and to some extent tunneling.  

We present some results from an approach
that uses semiclassical electron dynamics to evaluate the correlation
term in the propagation of the reduced density-matrix, with all the
other terms in the equation of motion treated exactly, as introduced in
Refs.~\onlinecite{RRM10,EM11}; thus it is a semiclassical-correlation driven
time-dependent Hartree-Fock (TDHF).  We study dynamics in perturbative
and non-perturbative fields in two one-dimensional ($1d$) model systems
of two electrons: one is a Hooke's atom, and the other a soft-Coulomb
Helium atom. The method improves over both TDHF and the pure
semiclassical method for the dynamics and excitation
spectra in the Hooke's atom case, but gives unphysical negative density regions in the
soft-Coulomb case. This is due to violation of $N$-representability
conditions, even though the pure semiclassical dynamics and the TDHF
on their own preserve these conditions.

\section*{Propagating Reduced Density Matrices: A Brief Review}
\label{sec:review}
We start with the time-dependent Schr\"odinger equation for the electron dynamics of a given system (defined by the external potential, $v\ext(\br)$): 
\ben
i\dot\Psi(x_1..x_N,t) = \left( \sum_j -\nabla_j^2/2 + \sum_j v\ext(\br_j) + \sum_{i<j}v\inter(\br_i,\br_j) \right)\Psi(x_1..x_N,t)
\label{tdse}
\een
where $x=(\br,\sigma)$ is a combined spatial and spin index, $\Psi$ is
the wavefunction, and $v\inter$ is the $2$-body Coulomb interaction
between the electrons, $v\inter(\br,\br') = 1/\vert\br-\br'\vert$. Atomic units are used
throughout this paper, ($m_e = \hbar = e^2 = 1$). Additionally the initial wavefunction $\Psi_0$ must 
be specified in order to begin the propagation. However solving
Eq. (\ref{tdse}) is computationally an extremely costly exercise and
becomes intractable as the number of electrons in the system
grows. Thus we must seek an alternative approach that aims to
reproduce the result of Eq. (\ref{tdse}) but at a much more reasonable
computational cost. Further, as mentioned in the introduction, the many-electron wavefunction contains far more information than one usually needs. 
 
Most observables that are experimentally measurable or of interest involve one- and two-body operators, such as the density e.g. in  dipole/quadrupole moments, the momentum-density e.g. in Compton profiles, and pair-correlation functions in e.g. double-ionization. So a formulation directly in terms of one- and two-body density-matrices, bypassing the need to compute the many-electron wavefunction, would be more useful.

This leads to the concept of reduced density-matrices, where the $p$-RDM involves tracing the full $N$-electron wavefunction over $N-p$ degrees of freedom:
\ben
\rho_{p}(x'_1..x'_p, x_1...x_p, t) = \frac{N!}{(N-p)!} \int dx_{p+1}...dx_N \Psi^*(x'_1..x'_p,x_{p+1}...x_N,t)\Psi(x_1..x_p,x_{p+1}...x_N,t)
\label{pRDM}
\een
The diagonal of the $p$-RDM gives the $p$-body density, $\Gamma(x_1..x_p,t)$, the probability of finding any $p$ electrons at points $\br_1..\br_p$ with spin $\sigma_1..\sigma_p$ at time $t$.  
One can also define spin-summed RDMs: e.g. $\rho_1(\br',\br,t) = \sum_{\sigma, \sigma_2...\sigma_N}\rho_1(x',x,t)$. 
The  Bogoliubov-Born-Green-Kirkwood-Yvon (BBGKY) hierarchy of equations of motion for
the RDMs were written down sixty years
ago~\cite{BBGKY,Bonitz}: The first in the hierarchy is that for $\rho_1$, which spin-summed is:
\ben
i\dot{\rho_1}({\bf{r'}},{\bf{r}},t) = \left(-\nabla^2/2 + v\ext({\bf{r}},t)+ \nabla'^2/2 - v\ext({\bf{r'}},t)\right)\rho_1({\bf{r'}},{\bf{r}},t) + \int d^3r_2 f\ee(\br,\br',\br_2) \rho_2(\br',\br_2,\br,\br_2,t)
\label{rho1dot}
\een
where
\ben
f\ee(\br,\br',\br_2) = v\inter(\br,\br_2)-v\inter(\br',\br_2)
\een
The electron-interaction term in the equation for
the 1RDM involves the 2RDM, whose equation of motion, the second in the hierarchy,
\bea
\nonumber
i\dot{\rho_2}(\br'_1,\br'_2,\br_1,\br_2,t) = \bigg(\sum_{i=1,2}\Big(\left(-\nabla_i^2/2 + v\ext({\bf{r_i}},t)\right) - \left(-\nabla_i'^2/2 + v\ext({\bf{r_i'}},t)\right)\Big)+v\inter(\br_1,\br_2)\\
-v\inter(\br_1',\br_2')\bigg)\rho_2(\br'_1,\br'_2,\br_1,\br_2,t) + \int dr_3\Big(f\ee(\br_1,\br_1',\br_3)+f\ee(\br_2,\br_2',\br_3)\Big)\rho_3(\br_1,\br_2,\br_3,\br_1',\br_2',\br_3,t)
\eea 
involves the 3RDM, and so on.  Solving the full hierarchy is
equivalent to solving Eq.~(\ref{tdse}) and no less impractical for
many-electron systems. 
 The hierarchy is usually therefore truncated,
typically using a ``cluster expansion'' where one reconstructs
higher-order RDMs as antisymmetrized products of lower order ones plus
a correlation term, sometimes referred to as a cumulant. 
Putting the correlation term to zero becomes
exact for the case when the underlying wavefunction is a single Slater
determinant (SSD).  For example, in the case of truncation at the
first equation, the equations reduce to TDHF,  and, for a spin singlet of a
closed shell system, $\rho_2$ in Eq.~(\ref{rho1dot}) is replaced by,
\ben
\rho_2^{\rm SSD}(\br',\br_2,\br,\br_2,t)=\rho(\br_2,t)\rho_1(\br',\br,t)-\frac{1}{2}\rho_1(\br',\br_2,t)\rho_1(\br_2,\br,t)
\label{rho2SSD}
\een

If instead, the truncation is done at the second equation in the BBGKY heirarchy, putting the
correlation term to zero in $\rho_3$, one obtains the Wang and Cassing
approximation~\cite{WC85,CVV93}.  One has then the choice of propagating the equation for $\rho_2$ alone  or propagating it simultaneously alongside the equation for $\rho_1$~\cite{AHNRL12}. 
However,
recently it was found that, in the former case, propagating while neglecting the
three-particle correlation term leads to the eventual violation of 
energy conservation~\cite{LBSI15,AHNRL12}.
Now the $p$-RDM can be obtained from higher-order RDMs via
contraction (i.e. partial trace),
\ben
\rho_p(x_1'..x_p', x_1..x_p,t) = \frac{1}{N-p}\int dx_{p+1} \rho_{p+1}(x_1'..x_p',x_{p+1}, x_1..x_p,x_{p+1}, t),
\label{contcons}
\een
as follows from the definition Eq.~(\ref{pRDM}). So an important condition
to consider when formulating reconstructions is whether they are
contraction-consistent, i.e. whether they satisfy Eq.~(\ref{contcons}).
In fact, in Ref.~\onlinecite{LBSI15}, it was shown that the
reconstruction approximation of Refs.~\onlinecite{WC85,CVV93} violated this condition; the underlying reason was that the
spin-decomposed three-particle cumulant\cite{Mazz98}, neglected in this approximation, has non-zero contraction.
 This
realization enabled the authors of Ref.~\onlinecite{LBSI15} to derive a
``contraction-consistent'' reconstruction for $\rho_3$ by including the part
of the three-particle cumulant that has non-zero contraction, which fortunately
is exactly known as a functional of the 2RDM. This was able to conserve  energy in the dynamical
simulations. However one cannot breathe easy just yet: propagation
with this contraction-consistent reconstruction violated
$N$-representability, another fundamental set of conditions that
wreak havoc if not satisfied.
Ref.~\onlinecite{AHNRL12} showed that contraction-consistency, and energy conservation, can be enforced if both 1RDM and 2RDM are propagated simultaneously, even while neglecting the three-particle correlation term.
However again, $N$-representability  was violated in this approach, leading to unphysical dynamics, instabilities and regions of negative densities. 

$N$-representability means that there exists an underlying
many-electron wavefunction whose contraction via Eq.~(\ref{pRDM}) yields
the matrix in question~\cite{C63,Mazz12}. For 1RDMs, the
$N$-representability conditions are simple, and usually expressed in terms of its eigenvalues $\eta_j$, called natural orbital (NO) occupation numbers, as defined via:
\ben 
\rho_1(\br',\br,t) = \sum_{j} \eta_j\xi^*_{j}(\br',t)\xi_{j}(\br,t) 
\label{NOexp}
\een
for the spin-summed singlet case,
where $\xi_{j}(\br,t)$ are natural orbitals. The $N$-representability conditions are  that $0\le \eta_j \le 2$, and $\sum_j \eta_j = N$.  (For the spin-resolved case, the first condition becomes $0\le \eta_j \le 1$). 
The 1RDM should be positive semi-definite, with trace equal N, and each eigenvalue bounded above by 2.  If
this is violated, densities can become negative in places, even when
the norm is conserved. (Note that it can be shown that particle number is always conserved by any approximation~\cite{AHNRL12}). 
  For 2RDMs, only very recently
has a complete set of conditions for ensemble $N$-representability
been discovered~\cite{Mazz12}; for pure states, not all the conditions are known, although some are. One
important condition regards positive semi-definiteness of the 2RDM, which is
challenging to maintain in dynamics when using approximate reconstructions. Yet without positive semi-definiteness, the propagation
becomes unstable.  The condition is in fact violated even by the
contraction-consistent reconstruction introduced in Ref~\onlinecite{LBSI15} and by the  joint 1RDM and 2RDM propagation in Ref.~\onlinecite{AHNRL12}.
Even for ground-state problems (where the analog of the BBGKY
equations is referred to as the contracted Schr\"odinger
equation), the
reconstruction functionals can violate such conditions, and iterative ``purification'' schemes
have been introduced to yield self-consistent $N$-representable
ground-state solutions~\cite{Mazz02,ACTPV05}. Even when the initial RDM satisfies
$N$-representability conditions, one can find violations building up at
later times in the propagation when approximate reconstruction
functionals are used~\cite{AHNRL12,LBSI15}. Ref.~\onlinecite{LBSI15} presented promising results
where a dynamical purification scheme was applied at each time-step in
the dynamics of molecules in strong-fields, leading to stable and
accurate propagation. A similar method~\cite{JD14} uses an energy-optimization procedure to obtain the 2RDM at each time-step while also enforcing various $N$-representability conditions, but the resulting dynamics is unable to change occupation numbers. 

From a different angle, it has been recently shown that the BBGKY equations can be recast into a Hamiltonian formulation~\cite{R12}, that opens the possibility of using advanced approximate methods of classical mechanics to analyze the equations and derive different reconstructions, in terms of equivalent classical variables.

On the other hand, ``time-dependent density-matrix functional
theory''(TDDMFT)~\cite{PGB07}, which deals only with Eq.~(\ref{rho1dot}),
proceeds from a somewhat different philosophy: the idea is that the
2RDM and all observables can in principle be obtained {\it exactly}
from the time-dependent 1RDM due to the Runge-Gross theorem of
TDDFT. The latter theorem~\cite{RG84,TDDFTbook2} proves that given an
initial state, there is a one-to-one mapping between the time-evolving
one-body density ($\rho(\br,t)$, diagonal of the 1RDM), and the
externally applied potential. This means that, in principle, knowledge
of $\rho(\br,t)$ is enough to determine the many-electron wavefunction,
up to a purely time-dependent phase, and hence all pRDMs also. Since
$\rho_1(\br,\br,t) = \rho(\br,t)$, this means in turn that
$\rho_1(\br',\br,t)$ determines all properties of the system. The only
assumption is that time-evolution of  $\rho_1$ occurs  in a local potential,
meaning a multiplicative operator in space, which raises questions about  $v$-representability~\cite{G15}.

There has been significant effort to approximate the 2RDM of
Eq.~(\ref{rho1dot}) as a functional of $\rho_1$, or of its NOs and occupation numbers (Eq.~(\ref{NOexp})). A natural starting
point is to insert the time-evolving 1RDM into an approximation
developed for ground-state density-matrix functional
theory~\cite{Mueller84,GU98,BB02,GPB05,LeivaPiris05,SDLG08}, thus
making an ``adiabatic'' approximation. These functionals can give very
good approximations for ground-state properties, especially important
for strongly-correlated systems where common approximations in alternative scalable methods
like density-functional theory fail. However, when used in
time-propagation, these same functionals keep the occupation numbers
fixed~\cite{GBG08,RP10,RP11,AG10,GGB10}, which leads to erroneous
dynamics. The first real-time non-perturbative application of TDDMFT~\cite{RP10} resorted to an extra energy-minimizing procedure to determine occupation numbers at each time-step that resulted in time-evolving occupation numbers. By considering perturbations around the ground-state, a
frequency-dependent response theory can be
formalized~\cite{GBG08} from which excitation energies can be
computed, and it was shown that adiabatic functionals cannot capture
double-excitations. Phase-including NO (PINO) functional theory~\cite{GGB10,GGB12,MGGB13} has been introduced to overcome this problem. Here the functional depends on the phase of the NO, which extends out of the realm of TDDMFT since 
any phase-dependence of the NOs cancels out when $\rho_1$ is formed.

Computationally, it has been argued that there is an advantage to
propagating the NOs and occupation numbers directly
instead of working with Eq.~(\ref{rho1dot})~\cite{AG10,BB13}.
By renormalizing the NOs via their occupation numbers,
$\tilde\xi(t)\rangle = \sqrt{\eta_i(t)}\vert \xi(t)\rangle$,
Refs.~\onlinecite{BB13, BRB14, RBB14} showed that the equations of motion
for the orbitals and those for the  occupation numbers can be instead combined into
one for each renormalized-orbital, which is numerically far more
stable than the coupled equations for $\eta_i(t)$ and $\vert\xi_i(t)\rangle$.  By
studying model two-electron systems, for which the exact 2RDM is known
in terms of the NOs and occupation
numbers~\cite{LS56}, Refs.~\onlinecite{BB13, BRB14, RBB14} could
propagate the renormalized NOs in strong
fields, with the only approximation being propagating a finite number
of orbitals. Even relatively few orbitals gave very good results for
challenging phenomena in correlated strong-field dynamics: autoionization, Rabi
oscillations, and non-sequential double-ionization. For more than two
electrons, one will however run again into the challenge of finding an
accurate approximation for the 2RDM in terms of the renormalized
NOs.

The progress and applications in time-propagating RDMs  as described above
has been relatively recent (the use of RDMs in static electronic
structure theory is much older and more established), although the
BBGKY equations were written down sixty years ago. This is partly
because of the instabilities stemming from violating
$N$-representability when one truncates the hierarchy, or the
inability of the adiabatic approximations for the functionals
$\rho_2[\rho_1]$ to change occupation numbers, as reviewed above.  TDDFT, on the
other hand, formulated about thirty years ago~\cite{RG84,TDDFTbook2},
has seen significant applications, especially for the calculations of
excitations and response, while the past decade has witnessed exciting
explorations into strong-field regime.  As discussed above, the
Runge-Gross states that all observables can be obtained from the
one-body density, but, instead of working directly with the density,
TDDFT operates by propagating a system of non-interacting electrons,
the Kohn-Sham system, that reproduces the exact interacting
density. The potential in the equation for the Kohn-Sham orbitals is
defined such that $N$ non-interacting electrons evolving in it have
the same time-dependent one-body density as the true interacting
problem. One component of this potential is the exchange-correlation
potential, a functional of the density $\rho$, the initial interacting state $\Psi$,
and the initial choice of Kohn-Sham orbitals $\Phi$ in which to begin the
propagation, $v\xc[\rho; \Psi_0, \Phi_0](\br, t)$. In almost all of the
real-time non-perturbative calculations, an adiabatic approximation is
used, in which the time-evolving density is inserted in a ground-state
functional approximation, neglecting the dependence on the
initial-states and the history of the density. This has produced
usefully accurate results in a range of situations, e.g.  modeling
charge-transfer dynamics in photovoltaic candidates~\cite{Carlo},
ultrafast demagnetization in solids~\cite{peter}, dynamics of molecules in strong fields~\cite{Bocharova}. Yet, there are errors, sometimes quite
large~\cite{RB09,RN11,RN12,RN12b,HTPI14}, and investigation of the behavior of the {\it
  exact} exchange-correlation potential reveals non-adiabatic features
that are missing in the approximations in use
today~\cite{EFRM12,FERM13,RG12}.  Further, when one is
interested in observables that are not directly related to the
density, additional ``observable-functionals'' are need to extract the
information from the Kohn-Sham system: simply evaluating the
usual operators on the Kohn-Sham wavefunction is not correct, even
when the exact exchange-correlation potential functional is
used~\cite{WB06,WB07,RHCM09,Henkel09}.  Although it is in principle possible to extract all observables from the Kohn-Sham system, it is not known how. A final challenge is that Kohn-Sham evolution maintains constant occupation numbers, even with the exact functional, which results in strong exchange-correlation effects. 
The one-body nature of the Kohn-Sham
potential means that the Kohn-Sham state remains a SSD throughout the
evolution, even though the interacting system that it is modeling can
dramatically change occupation numbers~\cite{AG10,LFSEM14}, evolving far from an SSD (e.g. if a singlet single excitation gets appreciably populated during the dynamics).   This leads to large features in the exact exchange-correlation potential that are difficult to model accurately. 

So, although computationally attractive, one could argue that
operating via a non-interacting reference leads to a more difficult
task for functionals. This has motivated the revisiting of the 1RDM
dynamics in recent years as discussed above: any one-body observable
can be directly obtained from the 1RDM using the usual operators, and
one does not need the effective potential to ``translate'' from a
non-interacting system to an interacting system. This suggests the terms
in the equation that contain the many-body physics could be easier to
model.  As discussed in the previous paragraphs, the difficulty then is to
come up with approximations for the 2RDM that can change occupation
numbers and maintain $N$-representability of the 1RDM.  In this work,
we implement the idea first introduced in Ref.~\onlinecite{RRM10}, using a
semiclassical approximation for the correlation term in
Eq.~(\ref{rho1dot}).

\section*{Semiclassical-Correlation Driven TDHF}
\label{sec:formalism}

From now on we deal only with singlet states and consider only the spin-summed RDMs. 
We begin by extracting the correlation component of Eq.~(\ref{rho1dot}), by decomposing $\rho_2$ via an SSD-contribution from Eq.~(\ref{rho2SSD}), plus a correlation correction:
\ben
\rho_2(\br',\br_2,\br,\br_2,t) = \rho_2^{\rm SSD}(\br',\br_2,\br,\br_2,t) + \rho\2c(\br',\br_2,\br,\br_2,t)
\label{rho2}
\een
 Then the last term of Eq. (\ref{rho1dot}) can be written 
\ben
\int d^3r_2 f\ee(\br,\br',\br_2) \rho_2(\br',\br_2,\br,\br_2,t) = \left(v\H(\br,t)-v\H(\br',t)\right)\rho_1(\br',\br,t) + F\x(\br',\br,t) + v\2c(\br',\br,t)
\label{eepart}
\een
where
\ben
v\H(\br,t) = \int d^3r' \ \frac{\rho(\br',t)}{\vert\br-\br'\vert}
\een
is the familiar Hartree potential of DFT and 
\ben
\label{fock}
F\x(\br',\br,t) = -\frac{1}{2}\int d^3r_2 f\ee(\br,\br',\br_2) \rho_1(\br',\br_2,t)\rho_1(\br_2,\br,t)
\een
is the Fock exchange matrix. The final term of Eq. (\ref{eepart}) we refer to as the correlation potential:
\ben
v\2c(\br',\br,t) = \int d^3r_2 f\ee(\br,\br',\br_2)\rho\2c(\br',\br_2,\br,\br_2,t)
\label{v2c}
\een
Without $v\2c$, the propagation of Eq.~\ref{rho1dot} using the first two terms of Eq.~\ref{eepart}, reduces to TDHF. 
In the present work we will evaluate $v\2c$ via semiclassical Frozen Gaussian dynamics, so turn now to a short review of this. 

\subsection*{Frozen Gaussian Dynamics}
Semiclassical methods aim to approximate the solution of Eq.~(\ref{tdse}) via an expansion in $\hbar$; the zeroth order  recovers the classical limit while the first-order $O(\hbar)$ terms are referred to as the semiclassical limit. For propagation, a popular example is the Heller-Herman-Kluk-Kay (HHKK)  propagator~\cite{H81,HK84,KHD86,TW04,K05,M98,GX98,S81,V28} where 
the $N$-particle wavefunction at time $t$ as a function of the $3N$ coordinates, denoted ${\underline{\underline{\br}}}=\{ \br_1,...,\br_N\}$, is:
\ben
\label{frozg} 
\Psi^{\rm FG}({\underline{\underline{\br}}},t) = \int\frac{d\bq_0d\bp_0}{(2\pi\hbar)^N}\langle{\underline{\underline{\br}}}\vert\bq_t\bp_t\rangle C_{\bq,\bp,t} e^{iS_t/\hbar}\langle\bq_0\bp_0 \vert\Psi_0\rangle
\een
where $\{\bq_t,\bp_t\}$ are classical phase-space trajectories at time $t$ in $6N$-dimensional phase-space, starting from initial points $\{\bq_0,\bp_0\}$. In
Eq.~(\ref{frozg}), $\langle{\underline{\underline{\br}}}\vert\bq\bp\rangle$ denotes the
coherent state:
\ben
\langle{\underline{\underline{\br}}}\vert\bq\bp\rangle = \prod_{j=1}^{3N}\left(\frac{\gamma_j}{\pi}\right)^{1/4}e^{-\frac{\gamma_j}{2}(r_j-q_j)^2 + ip_j(r_j-q_j)/\hbar}
\een
where $\gamma_j$ is a chosen width parameter. $S_t$ is the
classical action along the trajectory $\{\bq_t,\bp_t\}$. Finally, each
trajectory in the integrand is weighted by a complex pre-factor based
on the monodromy (stability) matrix, $C_{\bq,\bp,t}$ which guarantees the solution is exact to first order in $\hbar$. Computing this pre-factor is the most time-consuming element in the integral, scaling cubically with the number of degrees of freedom.

When the prefactor in Eq.~(\ref{frozg}) is set to unity, HHKK reduces
to the simpler Frozen gaussian (FG) propagation~\cite{H81}, which is 
more efficient. As a consequence, it is no longer exact to order
$\hbar$ and the results are no longer independent of the choice of
width parameter $\gamma_j$, unlike in HHKK. For our calculations we
take $\gamma_j=1$. Neither the HHKK propagation nor FG are unitary;
typically we find the norm of the FG wavefunction decreases with time,
and so we must renormalize at every time-step. In previous
work\cite{EM11}, the FG dynamics of electrons was investigated and
found to give reasonable results for a number of different quantities
and systems. Some of these will be referred to in the Results presented here.

\subsection*{TDDMFG}

In this work we will implement the idea of Ref. \onlinecite{RRM10} whereby a FG propagation, running parallel to a propagation of the 1RDM, is used to construct $v\2c$, which is then used in Eq. (\ref{eepart}) and Eq. (\ref{rho1dot}) to propagate the 1RDM. 
From the FG wavefunction given by Eq. (\ref{frozg}), the 1RDM and 2RDM  are computed and then used to construct:
\ben
\label{v2cfg}
v\FG\2c(\br',\br, t) = \int d^3r_2 f\ee(\br,\br',\br_2)\rho\FG\2c(\br',\br_2,\br,\br_2,t)
\een
where $\rho\FG\2c$ is found by inverting Eq. (\ref{rho2}):
\ben
\rho\FG\2c(\br',\br_2,\br,\br_2,t)=\rho\FG_2(\br',\br_2,\br,\br_2,t)-\rho\FG(\br_2,t)\rho\FG_1(\br',\br,t)+\frac{1}{2}\rho\FG_1(\br',\br_2,t)\rho\FG_1(\br_2,\br,t) 
\een
We then insert this into Eq.~\ref{v2cfg}, and propagate Eq.~\ref{rho1dot} with the last term evaluated using Eq.~\ref{v2cfg}.
We refer to this coupled dynamics as TDDMFG, meaning time-dependent density-matrix propagation with frozen-gaussian correlation.  

The scheme of Ref.~\onlinecite{RRM10} takes advantage of the
``forward-backward'' nature of the propagation of the 2RDM (i.e there
is both a $\Psi(t)$ and a $\Psi^*(t)$), which leads to some
cancellation of the oscillatory phase for more than two electrons. We
also observe that the spatial permutation symmetry of the
initial-state is preserved during the evolution (since the Hamiltonian
is for identical particles, exchanging coordinate-momentum pairs of
two electrons does not change the action). We will study here
two-electron singlet states where the wavefunction is
spatially-symmetric under exchange of particles. In the FG
propagation, although the energy of each classical trajectory is
conserved in the absence of external fields, the energy of the FG
wavefunction constructed from these trajectories is not guaranteed to
be. As noted earlier, the norm is not conserved either and the
wavefunction must be renormalized at each time. Thus, in general it is
unlikely that energy will be conserved in the TDDMFG scheme.

\subsection*{Computational Details}

The phase-space integral in Eq. (\ref{frozg}) is performed using Monte
Carlo integration, with the distribution of $M$ initial phase-space
points weighted according to a simple gaussian initial
distribution. In principle this method scales as $\sqrt{M}$,
however the oscillatory phase from the action $S_t$ can make the FG
propagation difficult to converge and thus a large number of
trajectories are often needed. This, in turn, means that
parallelization of the numerical computation of Eq. (\ref{frozg}) is
needed. Fortunately the main task is ``embarrassingly parallel'' as
each classical trajectory can be calculated separately, however
construction in real space of the FG wavefunction, 2RDM, 1RDM, and Eq. (\ref{v2cfg}) is
time-consuming and also required parallelization (in a manner similar to
the Fock exchange matrix calculations discussed below). To avoid
performing these costly procedures at every time step, we used a
linear-interpolation of Eq. (\ref{v2cfg}) which only required its
construction every $D_V$ timesteps. We tested that the results were
converged with respect to $D_V$, finding accurate results for values
as high as $D_V=200$ for a timestep of $dt=0.001$ au for the cases we studied.

The 1RDM propagation was performed via the predictor-corrector method combined with an Euler forward-stepping algorithm. All quantities were calculated on a real-space grid and the derivatives in Eq. (\ref{rho1dot}) were done using a $3$-point finite difference rule. 

Calculation of the Fock exchange matrix, Eq. (\ref{fock}), also required parallelization, as it has the worst scaling (cubic) with respect to the number of grid points of the remaining quantities. Parallelization often contains additional subtleties which can make the problem non-trivial, thus in order to parallelize efficiently, the problem was first transformed to resemble a more typical problem in parallel computing. To detail this procedure, it is convenient to switch to a matrix representation:
\ben
\boldsymbol{\rho}\coloneqq\rho_{nm}=\rho_1(\br_n,\br_m)
\een
where $\br_n$ is the $n$\textsuperscript{th} point of the real-space grid. We then define a new matrix
\ben
\boldsymbol{\tilde{\rho}}\coloneqq\tilde{\rho}_{nm} = A_{nm}\rho_{nm}
\een
where
\ben
\mathbf{A}\coloneqq A_{nm} = v\inter(\br_n,\br_m)
\een
which is then used to construct 
\ben
\mathbf{C} = \boldsymbol{\rho}\boldsymbol{\tilde{\rho}}\coloneqq C_{nm} = \sum_k \rho_{nk}\tilde{\rho}_{km}
\een
The Fock exchange matrix can then be written as
\ben
\mathbf{F\x} = \frac{\Delta x}{2}\left(\mathbf{C}^\dagger - \mathbf{C}\right)  
\een
in the case when the Fock integral is evaluated via quadrature and $\Delta x$ is the grid spacing. Thus, we have reduced the calculation of the Fock exchange integral to the calculation of $\mathbf{C}$, which only involves a matrix-matrix multiplication. In parallel computing, matrix-matrix multiplication is a well-studied problem for which standard solutions exist and thus could be easily implemented in our code without additional difficulty.

\section*{RESULTS}
\label{sec:results}

In this section we present the results of the TDDMFG formulation for
various time-dependent problems and compare the results to the TDHF
(i.e. $v\2c=0$), the pure FG, and the exact cases. In order to compare
to exact results we work in $1d$ and focus on two-electron systems, as it allows us to solve the full TDSE in a reasonable time with reasonable
computational resources.

We first tested our TDDMFG propagation algorithm by coupling to the exact dynamics, i.e. we used the exact wavefunction to calculate the exact $v\2c$ at each time which is then used within the 1RDM propagation to verify it recovers the exact dynamics. To remove any error due to the initial ground state we start the FG dynamics in the exact GS wavefunction and the 1RDM propagation in the exact GS density matrix. 

Each of TDHF and FG calculations on their own can yield reasonably good results for particular cases, thus the goal of the coupled TDDMFG propagation should be to either improve upon both, or at least, improve the results in scenarios where one or the other performs poorly. 

\subsection*{Hooke's atom}

\begin{figure}[tb]
\unitlength1cm
\begin{picture}(12,6.2)
\put(-4.9,-3.6){\makebox(12,6.2){
\includegraphics{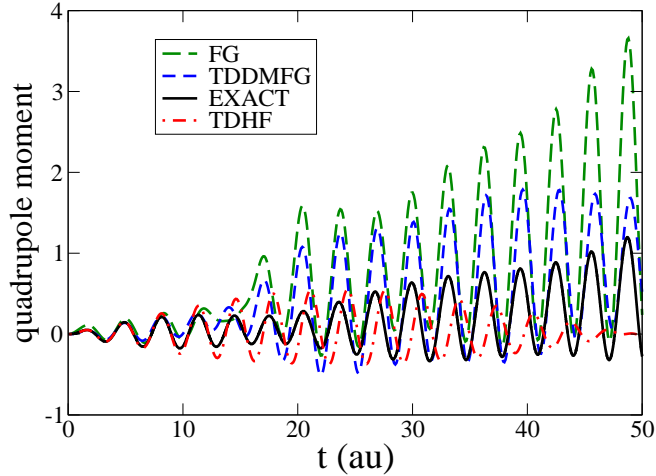}
}}
\end{picture}
\caption{Comparison of the quadrupole moment for Hooke's atom driven by a quadratic field at frequency $\omega=2$ for the exact case, TDHF, FG, and the coupled TDDMFG. For the FG run, a total of $100224$ classical trajectories were used.}
\label{f:hquad}
\end{figure}

We begin by studying Hooke's atom in $1d$, which consists of a harmonic external potential:
\ben
v\ext(x) = \frac{1}{2}k_0x^2
\een
and a softened Coulomb interaction between the electrons:
\ben
v\inter(x',x) = \frac{1}{\sqrt{(x-x')^2+1.0}}
\een
In our first application we drive the system by applying an oscillating quadrupole field 
\ben
\delta v\ext(x,t) = k(t)x^2
\een
where $k(t)=-0.025 \sin(2t)$ and $k_0=1$. The frequency of this perturbation is chosen to be resonant with an allowed excitation of the system. We then compare in Fig.~\ref{f:hquad} the change in the quadrupole moment relative to the ground state quadrupole ($Q_0$), 
\ben
Q(t) = \int dx \ \rho(x)x^2 - Q_0  
\een

as computed via exact propagation,  TDHF,  FG semiclassical
dynamics, and TDDMFG propagation. In this case we see an improvement to both the TDHF and
the FG calculations. While the exact quadrupole is seen to continue
increasing in amplitude over time, the TDHF does not, and in fact
oscillates in a beating pattern. This is partly due to the fact
that TDHF spectra cannot describe this particular excitation (which
will be discussed in more detail later) and leads to an off-resonance
Rabi oscillation. However even running TDHF at the  resonance frequency of TDHF does not show Rabi oscillations either; although the quadrupole begins to grow, it ultimately fails because of the spurious detuning effect explained in Ref.~\onlinecite{FLSM15}.   The FG, in
contrast, overestimates the amplitude of the oscillations, while the
TDDMFG coupled dynamics lies in between these two extremes and is much
closer to the exact result. In previous work we found that the
quadrupole is more sensitive than other quantities to the number of
trajectories used in our FG calculation, in this case $100224$. Thus
we could expect  better agreement if we further
increase the number of trajectories.  It is interesting to note that
all three approximate calculations work reasonably well for the first
$10$ au.

\begin{figure}[tb]
\unitlength1cm
\begin{picture}(12,6.2)
\put(-4.9,-3.6){\makebox(12,6.2){
\includegraphics{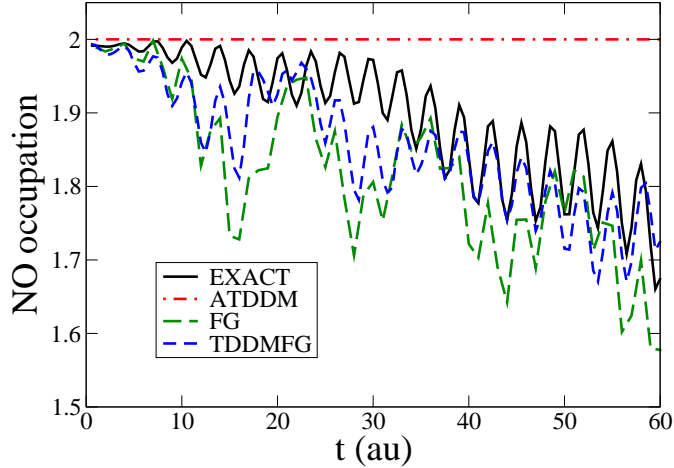}
}}
\end{picture}
\caption{Evolution of the occupation of the highest occupied natural orbital for the driven Hooke's atom in the exact case, TDHF, FG, and TDDMFG.}
\label{f:hNOs}
\end{figure}

Buoyed on by this success, we next examine the NO
occupation numbers to probe the 1RDM in more detail than 
the quadrupole, an averaged expectation value, can provide.
 As noted in the earlier review, a major shortcoming of adiabatic
functionals in TDDMFT is their inability to change these occupations. As
can be seen in Fig. \ref{f:hNOs} this behavior manifests itself as
straight lines, labelled ATDDM, constant at the initial state NO
occupations. This is also true for the TDHF case shown previously. Since we  spin-sum, the NOs go between 0 and 2,  and we plot the highest occupied NO occupation, which
this case begins very close to $2$, indicating the initial state is
strongly of an SSD character. The time-dependence
of this occupation is shown in Fig. \ref{f:hNOs}. The exact value
decreases towards a value of $1$ as the system becomes excited and the
wavefunction moves away from SSD-like
state. In Ref. \onlinecite{EM11}, it was shown that FG propagation can
quite accurately capture the evolution of the NO occupations, so the
question is whether the coupled dynamics of TDDMFG is also able to do
so. Examining the TDDMFG values, we see that TDHF coupled to the FG correlation
is  able to evolve the occupations accurately.
In fact TDDMFG is slightly better than the pure FG values
where FG has sometimes spuriously large oscillations. 

Although the amplitude of the oscillations of the TDDMFG quadrupole
are closer to those of the exact than the FG oscillations, the phase
of the oscillations of the latter is closer to the exact case than the
TDDMFG case. This becomes more evident carrying the propagation out to
longer times.  This can be understood from considering the resonant
frequencies of the system: the frequency of the perturbation $k(t)$ is
on-resonance with an excitation of the exact system which FG in fact
correctly describes. The TDDMFG excitation frequency is however
shifted slightly leading to the difference in phase observed here.  We
will now examine the respective excitation frequencies of each method
in more detail.

\begin{table}
\begin{tabular}{|c|c|c|c|}\hline
EXACT 	 & TDHF & FG  & TDDMFG	\\ \hline
1.000000 & 0.99 & 1.0 & 0.99	\\
1.734522 & 1.86	& 1.6 & 1.58	\\
2.000000 & - 	& 2.0 & 1.86	\\
2.734522 & 2.79	& 2.6 & 2.58	\\
3.000000 & -	& 3.0 & 2.77	\\ \hline
\end{tabular}
\caption{\label{t:hooke}The singlet excitation frequencies $\omega_n=E_n-E_0$, where the ground-state energy is $E_0 = 1.774040$ a.u., solved exactly for Hooke's atom, and the corresponding TDHF,FG, and TDDMFG values as calculated by real-time linear response.}
\end{table}

\begin{figure}[th]
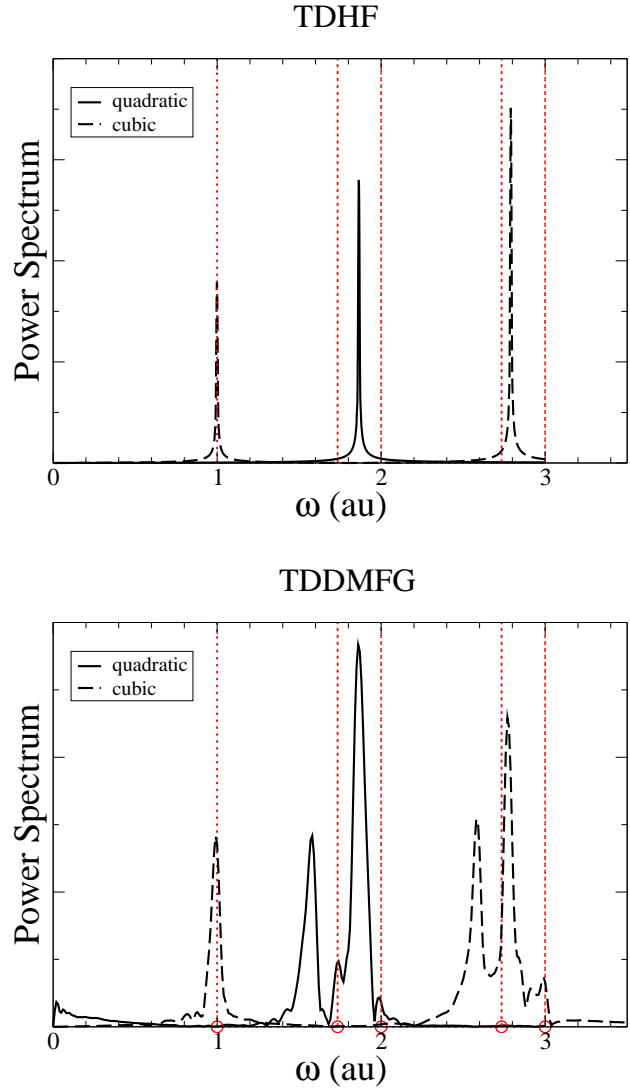

\unitlength1cm
\begin{picture}(12,15.1)
\put(-4.9,-3.4){\makebox(12,6.2){
\includegraphics{qpec_tddmfg.eps}
}}
\put(-4.9,4.1){\makebox(12,6.2){
\includegraphics{qspec_aexx.eps}
}}
\end{picture}
\caption{Excitation spectra of Hooke's atom calculated via linear response with quadratic/cubic kicked initial states and the Fourier transform of the relevant moment for TDHF (upper panel) and TDDMFG (lower panel).}
\label{f:hqspec}
\end{figure}

Turning to how well the TDDMFG captures excitation spectra, we focus
in particular on 
so-called double excitations (defined loosely as excitations which
have a large fraction of a doubly excited state character, with
respect to the SSD ground state of a
non-interacting reference system~\cite{EGCM11}).  It is known that these excitations
are missing from all TDDFT spectra calculations within the adiabatic
approximation. Since TDHF is equivalent to adiabatic exact exchange in TDDFT  for the
$2$-electron case studied here, we do not see double
excitations in the uncoupled 1RDM TDHF spectra. This point
is illustrated in the upper panel of Fig. \ref{f:hqspec} which shows
the TDHF power spectra. The spectra are  calculated via the linear response method,
utilizing a 'kicked' initial state defined as: \ben \Psi_0(x,y) =
e^{ik(x^n+y^n)}\Psi\GS(x,y) \een where $\Psi\GS$ is the ground-state
wavefunction, $k$ is a small constant, and $n$ is an integer (we
define a quadratic kick as $n=2$ and a cubic kick as $n=3$). An
expression for the initial 1RDM can be easily derived from this. A dipole kick ($n=1$) is commonly used when
calculating optical spectra as it corresponds to an electric field
consisting of a $\delta-$function at time $t=0$ which excites all dipole
allowed excitations~\cite{YNIB06}. However due to symmetries of Hooke's atom, to
access the double excitations, higher moment kicks were necessary~\cite{EM11}. To obtain the spectra, for each run we calculate the appropriate
moment (e.g. quadrupole moment for quadratic kicks) and Fourier
transform to frequency space to reveal the excitation peaks~\cite{EM11}. In the TDHF
case, we do not see the pair of excitations peaks at frequencies
(1.73,2.0), nor the pair (2.73,3.0) but instead see a single peak in
between. This behavior is commonly seen for TDDFT calculations with an
adiabatic approximation, where a frequency-dependent XC kernel is
required to split the peak into two separate excitations~\cite{MZCB04,EGCM11}; any adiabatic approximation in TDDMFT will also only display one peak~\cite{GBG08}.

Moving to the lower panel of Fig. \ref{f:hqspec}, we plot the TDDMFG
spectra calculated in the same manner. It can be seen that including
$v\FG\2c$ into the TDHF propagation correctly splits the single peak into two peaks, for both the quadratic and cubic
kick cases. Thus we have demonstrated that our coupled dynamics does
indeed capture double excitations. Identifying the
position of the peaks, we compare in Table \ref{t:hooke} the values
given by each method for the lowest 5 excitations of Hooke's atom. It
was found in Ref. \onlinecite{EM11} that FG on its own also describes
double excitations quite well, and in fact is exact for certain
excitations in Hooke's atom. This is due to the fact that the Hamiltonian 
becomes separable in 
center-of-mass and relative coordinates, and that in the center-of-mass coordinate is a simple harmonic oscillator~\cite{EM11}. 
It is
well-known that harmonic potentials are a special case for
semiclassical methods as they often perform exactly. With this in
mind, although the value of the TDDMFG frequencies are worse than the pure FG
values, they are competing with a special case, but in fact the
splitting between the peaks is  better described by the
TDDMFG. In particular, in the second multiplet, the exact splitting is 0.27, while that of the TDDMFG is 0.28, improved over the bare FG result of 0.4 (and obviously over TDHF where there is no peak). 
We emphasize again that these excitations are completely
missing in the TDHF case, or in any adiabatic TDDFT or TDDMFT functional.  It is better to have the
excitations shifted slightly from the exact result than to not
describe them at all.

\clearpage

\subsection*{Soft-Coulomb Helium}
We now move to the more realistic case of soft-Coulomb Helium where
the external potential is:
\ben
v\ext(x) = -\frac{2}{\sqrt{x^2+1.0}}
\een
which mimics the $3d$ case as for large $x$ it decays as $-1/|x|$. In the previous
case of Hooke's atom, we saw that the problem was well described by
the FG dynamics whereas the TDHF performed poorly (i.e. not
changing the NO occupations or capturing double excitations). Driving the TDHF with FG correlation  in TDDMFG interestingly improved over FG for the NO occupations and quadrupole moment, with slightly worse performance for the double-excitations.  For dynamics in the soft-Coulomb Helium 
case we will see, in contrast, that the FG is worse than the TDHF for some
quantities. Will the coupled
dynamics of TDDMFG improve the situation? As detailed in Ref. \onlinecite{EM11}, this case is much
more difficult for the FG method due to classically auto-ionizing trajectories (where one electron gains energy from the other and ionizes while the other slips below the zero-point-energy), thus a far greater number of trajectories are required: . Ref.~\onlinecite{EM11} discussed how in a truly converged Frozen Gaussian calculation, the contributions from these unphysical trajectories cancel each other out, but a very large number of trajectories are required; otherwise methods to cut out their contribution to the semiclassical integral can be used.
 In the
presented calculations, $2000448$ trajectories were used and all were kept. 

\begin{figure}[tb]
\unitlength1cm

\begin{picture}(16.15,8)
\put(-4.0,-4.6){
\makebox(7,7){\includegraphics{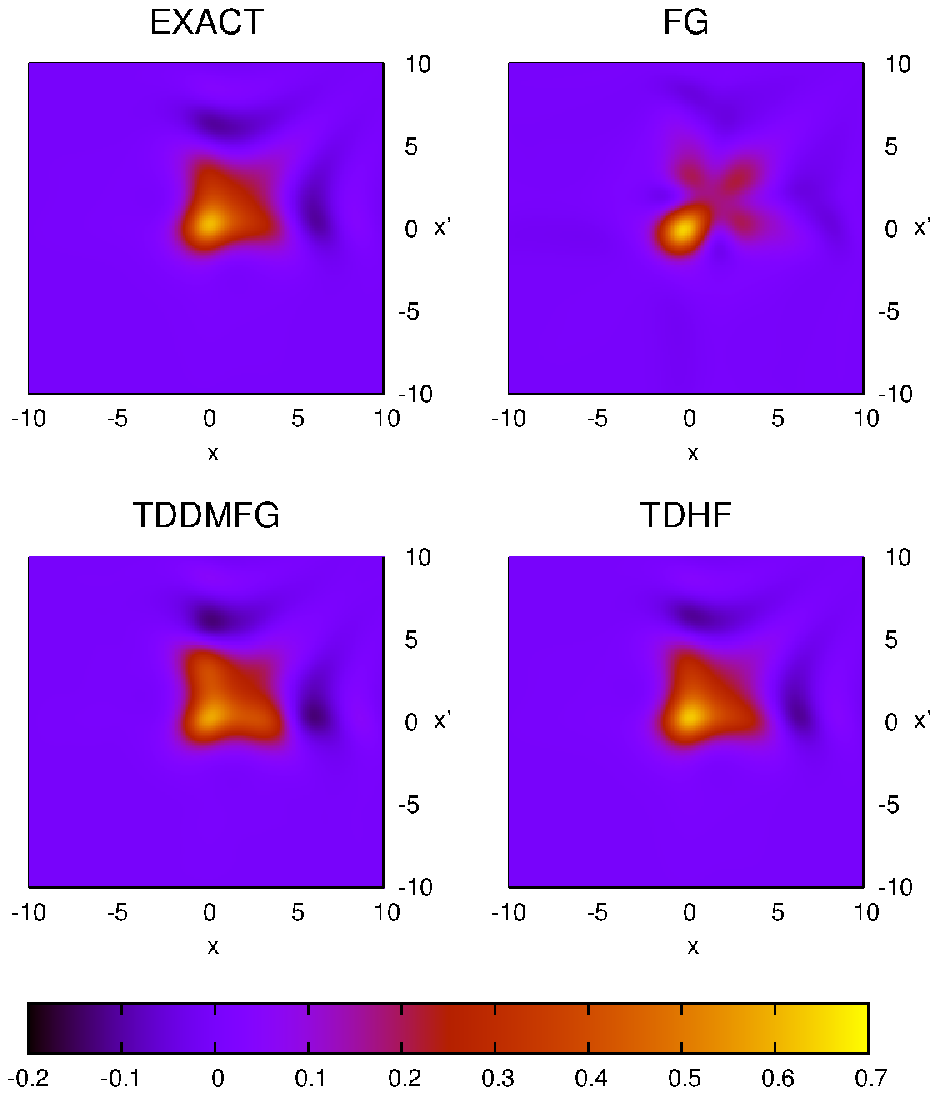}}}
\put(4.2,-4.6){
\makebox(7,7){\includegraphics{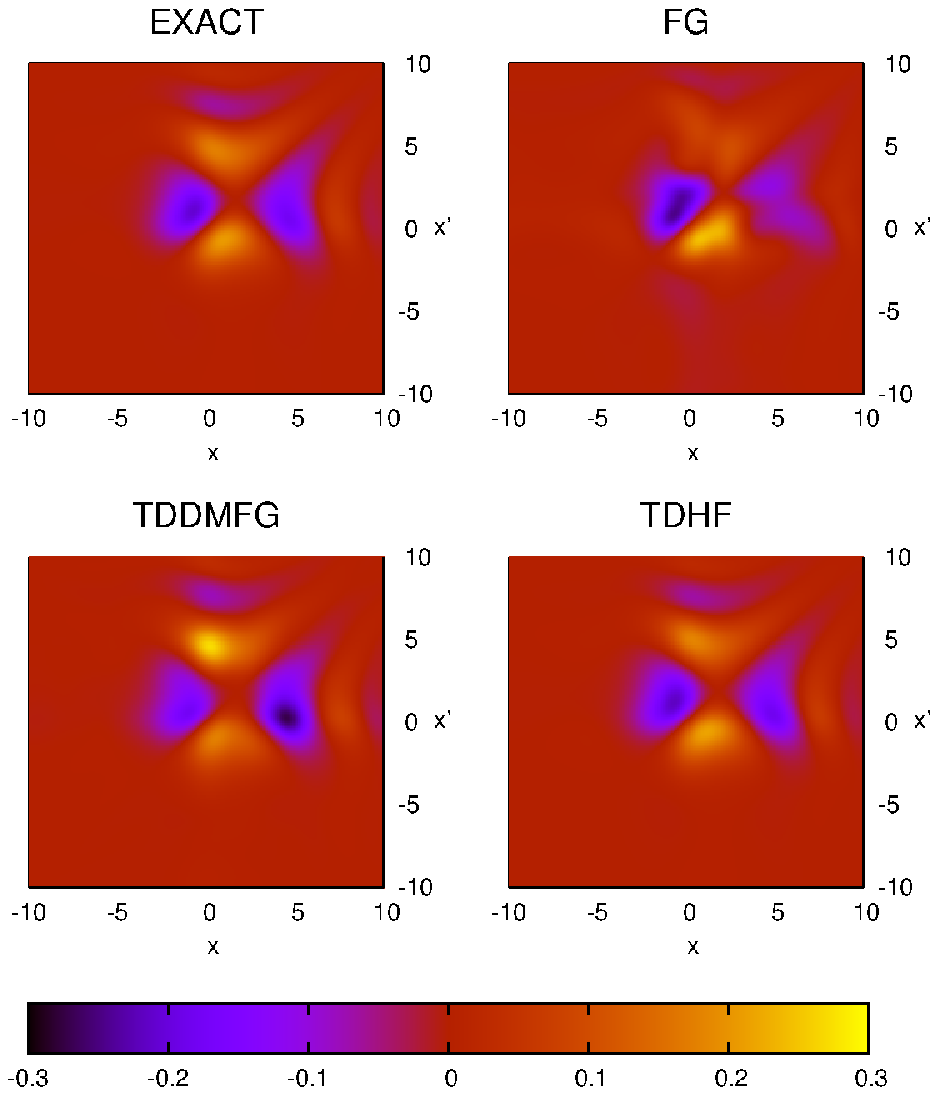}}}
\end{picture}
\caption{Comparison of the real (left panels) and imaginary (right panels) parts of the density matrix at time $T=10$ au for strongly driven soft-Coulomb Helium.}
\label{f:dmR10}
\end{figure}

We apply a strong laser pulse with a linearly-switched-on electric field:
\ben
\epsilon(t) = \frac{1}{\sqrt{2}}\cos(0.5t) \left\{\begin{array}{c l} \frac{t}{20} & t\leq 20 \\ 1 & t>20 \end{array}\right.
\label{eq:trapeps}
\een 
which is included in our Hamiltonian via the dipole approximation,
i.e. $\delta v\ext(x) = \epsilon(t)x $. We begin by examining the 1RDM
itself at time $T=10$ au: both the real and imaginary parts are shown in Fig. \ref{f:dmR10}. At this time, while the
structure of the FG 1RDM is broadly correct, it can be seen that the
TDHF 1RDM is much closer to the exact. The TDDMFG 1RDM also captures
more of the correct structure compared to the pure FG case, although
it generally overestimates the peaks and valleys. Thus, while the
$v\FG\2c$ is constructed from the poorer FG calculation, the TDDMFG
 follows more closely the more accurate TDHF
description.

\begin{figure}[tb]
\unitlength1cm
\begin{picture}(12,6.2)
\put(-4.9,-3.6){\makebox(12,6.2){
\includegraphics{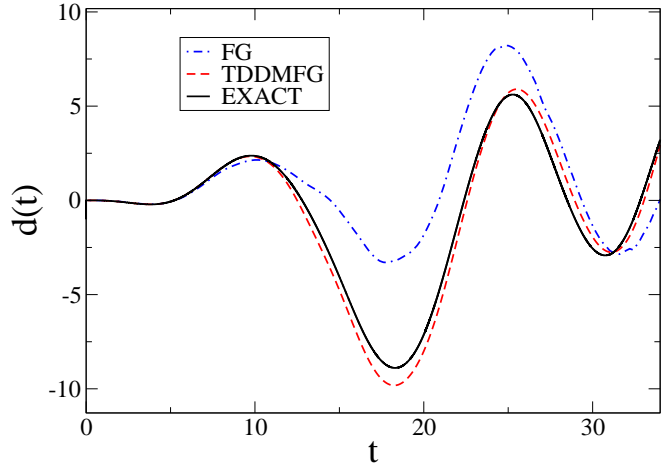}
}}
\end{picture}
\caption{The dipole moment during the strongly driven soft-Coulomb Helium for the exact calculation, the FG, and the TDDMFG. The FG calculation required $2000448$ classical trajectories. The TDHF calculation (not shown) is very accurate during this time period.}
\label{f:sdips}
\end{figure}

We next turn to the dipole moment which is plotted in
Fig. \ref{f:sdips}. The TDHF (not shown) essentially matches
the exact case, whereas the FG performs quite poorly, particularly
during the second optical cycle. The TDDMFG, in contrast, is
performing particularly well and follows very closely the exact
result, even at times greater than $T=10$ au, likely due to the guidance of the TDHF component in the evolution. 

At this point one might conclude  that the TDDMFG is behaving correctly,
however the good results for the 1RDM and dipole moment are masking
the fact that the underlying description suffers from a major error,
described below.

\begin{figure}[tb]
\unitlength1cm
\begin{picture}(12,6.2)
\put(-4.9,-3.6){\makebox(12,6.2){
\includegraphics{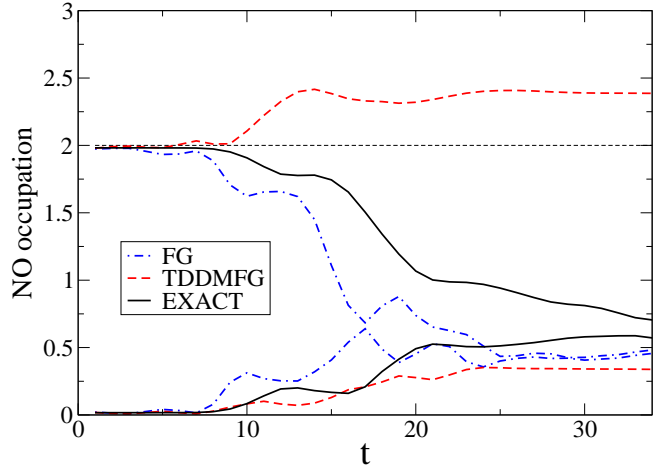}
}}
\end{picture}
\caption{The evolution of the two-highest occupied natural orbital occupations during the strongly driven soft-Coulomb Helium. We omit the ATDDM occupations as they stay constrant at their initial values.}
\label{f:sNOs}
\end{figure}

As was the case for Hooke's atom, a more thorough examination of the
method is given by studying the NO occupations. The
highest two NO occupations are plotted in Fig. \ref{f:sNOs} where the
strong field causes a large change in their values. In fact we see that
the FG description of the NO occupations is working better than we previously
anticipated, albeit
overestimating their change. Again the TDHF occupations (not shown) are constant,
fixed at their initial values. At $T=10$ au, the exact 1RDM is still
dominated by the highest occupied natural orbital, explaining why the
TDHF appeared so good at this time.

Unfortunately in the TDDMFG case, we see that the highest occupation
rises above $2$ thus violating the exact condition for
$N$-representability of the 1RDM (positive semidefiniteness).  The effect of this is
quite drastic as the density develops negative regions, due to having
negative occupations (the sum of the occupations remains $2$, thus an
increase above $2$ is accompanied by negative values). At $T=10$ au
the value of the highest NO is only slightly above $2$ and so this
bad behavior has yet to truly manifest itself.  The FG NO occupation
at $T=10$ au is underestimated, consistent with the FG 1RDM being not so accurate. At later times, none of the methods provide a particularly good
description of the 1RDM structure, with the TDHF remaining the closest, despite its constant occupation number, but the TDHF momentum densities are not good. 
The TDDMFG 1RDM resembles the TDHF but with exaggerated highs and
lows, and unphysical negative regions, manifest also in the momentum density. 

It is a frustrating situation where neither the FG and TDHF on their
own violate the N-representability condition, and FG does evolve the occupation numbers unlike any adiabatic approximation, 
however coupling FG correlation to TDHF in the
TDDMFG leads violation of $N$-representability. We speculate this is
due to a mismatch in correlation potential and the Hartree-exchange
terms as there is no mechanism to provide feedback between the two
calculations. Implementing a dynamical purification scheme along the
lines of that in Ref.~\onlinecite{LBSI15} that iteratively decreases the
magnitude of the negative occupation numbers should be investigated.

\section*{CONCLUSIONS}
\label{sec:Conclusions}
The TDDMFG dynamics which uses a Frozen Gaussian calculation
to construct an approximation to the correlation potential in 
 1RDM propagation gives mixed results. On one hand, it was
shown for Hooke's atom to be a significant improvement over all
adiabatic 1RDM functionals as it can vary the natural
orbital occupations reasonably accurately. Furthermore it was shown that double
excitations, which are difficult to capture with the commonly-used
TDDFT method and adiabatic TDDMFT, can  be accurately described with the TDDMFG
formalism. On the other hand, for soft-Coulomb Helium the method was
seen to fail drastically giving the unphysical result of negative
density due to violation of an exact constraint. Further work is
required to understand why this violation occurs and then hopefully to
then use this information to prevent it from occurring. 
This same problem, violation of the positive semi-definiteness, was also found in other recent approaches propagating RDMs with approximate correlation terms~\cite{AHNRL12,LBSI15}. 
Dynamical purification schemes along the line of that successfully used in Ref.~\onlinecite{LBSI15} could be very helpful here. Indeed
understanding how the FG correlation potential changes NO occupation
or includes double excitations could be used to construct better
approximations for TDDMFT. Finally, in this work we studied systems of only $2$
electrons, whereas we might expect that (semi)classical  methods work
best for large numbers of particles, and it remains to be seen whether
 the problems we encountered become less significant for larger systems.

An advantage of using a semiclassical scheme to evaluate the
correlation term, is that initial state dependence is automatically
taken care of: in general reconstructions, whether one begins in an
excited state or ground-state, the same approximation for the 2RDM in
terms of the 1RDM is assumed. This is known to be incorrect; several
different initial wavefunctions and different initial 2RDM's can give
rise to the same initial 1RDM~\cite{EM12,MB01}. The resulting correlation effect is
clearly different depending on the wavefunction, but this effect is
ignored in all reconstructions in use today. For this reason, it seems
worthwhile to pursue further investigations of a
semiclassical-correlation driven TDHF, once the $N$-representability
problem is taken care of, especially since in many simulations of
non-equilibrium dynamics of interest today, the problem starts in a
photo-excited state.

\subsection*{ACKNOWLEDGMENTS}         
We thank Robert Numrich and Richard Walsh, of the CUNY HPC center, for useful discussions concerning the parallelization of our code. PE acknowledges support by SFB 762 of the Deutsche Forschungsgemeinschaft. NM thanks NSF (Grant CHE-1162784) for financial support. The CUNY HPCC is operated by the College of Staten Island and funded, in part, by grants from the City of New York, State of New York, CUNY Research Foundation, and National Science Foundation Grants CNS-0958379, CNS-0855217 and ACI 1126113.

%\clearpage

\end{document}

%% file: dft.tex
\def\bea{\begin{eqnarray}}
\def\eea{\end{eqnarray}}
\def\ben{\begin{equation}}
\def\een{\end{equation}}
\def\benu{\begin{enumerate}}
\def\enu{\end{enumerate}}
\def\bei{\begin{itemize}}
\def\eei{\end{itemize}}

\def\sss{\scriptscriptstyle\rm}

\def\1var{(\bx_1...\bx\N)}

\def\bp{{\bf p}}
\def\br{{\bf r}}

\def\bx{{\br t}}

\def\bq{{\bf q}}

\def\x{_{\sss X}}

\def\xc{_{\sss XC}}

\def\N{_{\sss N}}
\def\H{_{\sss H}}

\def\ext{_{\rm ext}}

\def\ee{_{\rm ee}}

\def\sph_int{ {\int d^3 r}}

%% file: p16_02_11.bbl
\begin{thebibliography}{99}

\bibitem{RDMbook}
{\it Advances in Chemical Physics {\bf 134}: Reduced-Density-Matrix Mechanics: With Application to Many-Electron Atoms and Molecules}, Ed. D. A. Mazziotti, (John Wiley and Sons, Hoboken NJ, 2007)

\bibitem{EFB07}  
P. Elliott, F. Furche, and K. Burke, in Reviews in Computational
Chemistry, edited by K. B. Lipkowitz and T. R. Cundari (Wiley,
Hoboken, NJ, 2009), p. 91.

\bibitem{TDDFTbook2}
{\it Fundamentals of Time-Dependent Density Functional Theory}, Eds. M. A. L. Marques, N. T. Maitra, F. Nogueira,  A. Rubio, E.K.U. Gross, (Springer-Verlag, Berlin, 2012). 

\bibitem{PG15}
{\it Reduced Density Matrix Functional Theory (RDMFT) and Linear Response Time-Dependent RDMFT (TD-RDMFT)}, 
K. Pernal and K. J. H. Giesbertz, in {\it Density Functional Methods for Excited States}, Vol. 368 of Topics in Curr. Chem. (2015). 

\bibitem{V28}
J.H. van Vleck, Proc. Natl. Acad. Sci. USA {\bf 14}, 178 (1928).

\bibitem{S81}
{\it Techniques and Applications of Path Integration},L. S. Schulman, (Wiley \& Sons, Inc., 1981).

\bibitem{H81}
E. J. Heller, J. Chem. Phys. {\bf 75}, 2923 (1981). 

\bibitem{M98}
%{\em Quantum and Semiclassical Theory of Chemical Reaction Rates},
W. H. Miller, 
 Faraday Disc. Chem. Soc. {\bf 110}, 1 (1998).

\bibitem{TW04}
%{\em Semiclassical Description of Molecular Dynamics Based on Initial-Value Representation Methods},
M. Thoss and H. Wang, 
Ann. Rev. Phys. Chem. {\bf 55} , 299 (2004). 

\bibitem{K05}
K. G. Kay, Annu. Rev. Phys. Chem. {\bf 56} 255 (2005).

\bibitem{RRM10}
 A.K. Rajam, I. Raczkowska, N.T. Maitra,
%{\it  Semiclassical electron correlation in density-matrix propagation}, 
Phys. Rev. Lett. {\bf 105}, 113002 (2010).

\bibitem{EM11}
P. Elliott and N.T. Maitra, J. Chem. Phys. {\bf 135}, 104110 (2011).

\bibitem{BBGKY}
N. N. Bogoliubov, J. Phys. USSR {\bf 10}, 265 (1946); N. N. Bogoliubov and K. P. Gurov, J. Expt and Theor. Physics {\bf 17}, 614 (1947); J. Yvon, Actual. Sci.  Indust.{\bf 203} (Paris, Hermann, 1935); J. G. Kirkwood, J. Chem. Phys. {\bf 14}, 180 (1946), ibid {\bf 15}, 72 (1947); M. Born and H. S. Green, Proc. Roy. Soc. A. {\bf 188}, 10 (1946). 

\bibitem{Bonitz}
M. Bonitz, {\it Quantum Kinetic Theory} (B. G. Teubner, Stuttgart
Leipzig, 1998).

\bibitem{WC85}
S. J. Wang and W. Cassing, Ann. Phys. NY {\bf 159}, 328 (1985). 

\bibitem{CVV93}
F. Colmenero, C. P\'erez del Valle, and C. Valdemoro, Phys. Rev. A. {\bf 47}, 971 (1993). 

\bibitem{AHNRL12}
A. Akbari, M.J. Hashemi, A. Rubio, R. M. Nieminen, and R. van Leeuwen, Phys. Rev. B. {\bf 85}, 235121 (2012).

\bibitem{LBSI15}
F. Lackner, I. Brezinov\'a, T. Sato, K. L. Ishikawa, J. Burgd\"orfer, Phys. Rev. A, {\bf 91}, 023412 (2015). 

\bibitem{Mazz98}
D. A. Mazziotti, Chem. Phys. Lett. {\bf 289}, 419 (1998).

\bibitem{C63}
A. J. Coleman, Rev. Mod. Phys. {\bf 35}, 668 (1963). 

\bibitem{Mazz12}
D. A. Mazziotti, Phys. Rev. Lett. {\bf 108}, 263002 (2012). 

\bibitem{Mazz02}
D. A. Mazziotti, Phys. Rev. E {\bf 65}, 026704 (2002). 

\bibitem{ACTPV05}
D. R. Alcoba, F. J. Casquero, L. M. Tel, E. Prez-Romero, and C. Valdemoro, Int. J. Quantum Chem. {\bf 102}, 620 (2005). 

\bibitem{JD14}
D. B. Jeffcoat and A. E. DePrince, J. Chem. Phys. {\bf 141}, 214104 (2014).

\bibitem{R12}
R. Requist, Phys. Rev. A. {\bf 86}, 022117 (2012). 

\bibitem{PGB07}
K. Pernal, O. Gritsenko, and E. J. Baerends, Phys. Rev. A. {\bf 75}, 012506 (2007).

\bibitem{RG84}
E. Runge and E. K. U. Gross, Phys. Rev. Lett. {\bf 52}, 997 (1984). 

\bibitem{G15}
K. Giesbertz, J. Chem. Phys. Phys. {\bf 143}, 1 (2015). 

\bibitem{Mueller84}
A. M. K. M\"uller, Phys. Lett. A. {\bf 105}, 446 (1984). 

\bibitem{GU98}
S. Goedecker and C. J. Umrigar, Phys. Rev. Lett. {\bf 81},  866 (1998). 

\bibitem{BB02}
M. Buijse and E. J. Baerends, Mol. Phys. {\bf 100}, 401 (2002). 

\bibitem{GPB05}
O. V. Gritsenko, K. Pernal, and E. J. Baerends, J. Chem. Phys. {\bf 122}, 204102 (2005). 


\bibitem{LeivaPiris05}
P. Levia and M. Piris, J. Chem. Phys. {\bf 123}, 214102 (2005). 

\bibitem{SDLG08}
S. Sharma, J. K. Dewhurst, N. N. Lathiotakis, and E. K. U. Gross, Phys. Rev. B. {\bf 78}, 201103 (2008). 

\bibitem{GBG08}
K. Giesbertz, E.J. Baerends, O. Gritsenko, Phys. Rev. Lett. {\bf 101}, 033004 (2008).

\bibitem{RP10}
R. Requist and O. Pankratov, Phys. Rev. A {\bf 81}, 042519 (2010). 
\bibitem{RP11}
R. Requist and O. Pankratov, 
%{\it Time-dependent occupation numbers in reduced-density-matrix-functional theory: Application to an interacting Landau-Zener model}
Phys. Rev. A {\bf 83}, 052510 (2011).


\bibitem{AG10}
H. Appel and E. K. U. Gross, Europhys. Lett. {\bf 92}, 23001 (2010). 

\bibitem{GGB10}
K.J.H. Giesbertz, O. V. Gritsenko, and E. J. Baerends, Phys. Rev. Lett. {\bf 105}, 013002 (2010). 

\bibitem{GGB12}
K.J.H. Giesbertz, O. V. Gritsenko, and E. J. Baerends,
J. Chem. Phys. {\bf 136}, 094104 (2012).

\bibitem{MGGB13}
R. van Meer, O. V. Gritsenko, K.J.H. Giesbertz, and E. J. Baerends, 
J. Chem. Phys. {\bf 138}, 094114 (2013)

\bibitem{BB13}
M. Brics and D. Bauer, Phys. Rev. A {\bf 88}, 052514 (2013).

\bibitem{RBB14}
J. Rapp, M. Brics, and D. Bauer, Phys. Rev. A {\bf 90}, 012518 (2014).

\bibitem{BRB14}
M. Brics, J. Rapp, D. Bauer, Phys. Rev. A. {\bf 90}, 053418 (2014).


\bibitem{LS56}
P.-O. L\"owdin and H. Shull, Phys. Rev. {\bf 101}, 1730 (1956). 


\bibitem{Carlo}
C. A. Rozzi et al. Nat. Commun. {\bf 4}, 1602 (2013).

\bibitem{peter}
K. Krieger, J. K. Dewhurst, P. Elliott, S. Sharma, and E.K.U. Gross, J. Chem. Theory and Comput. {\bf 11}, 4870 (2015). 

\bibitem{Bocharova}
I. Bocharova et al., Phys. Rev. Lett. {\bf 107}, 063201 (2011).





\bibitem{RB09}
M. Ruggenthaler and D. Bauer, Phys. Rev. Lett. {\bf 102},
233001 (2009).
\bibitem{RN11}
S. Raghunathan and M. Nest, J. Chem. Theory Comput. {\bf 7},
2492 (2011).
\bibitem{RN12}
R. Ramakrishnan and M. Nest, Phys. Rev. A {\bf 85}, 054501
(2012).
\bibitem{RN12b}
S. Raghunathan and M. Nest, J. Chem. Theory Comput. {\bf 8},
806 (2012).
\bibitem{HTPI14}
B. F. Habenicht, N. P. Tani, M. R. Provorse, and C. M.
Isborn, J. Chem. Phys. {\bf 141}, 184112 (2014).

\bibitem{EFRM12}
P. Elliott, J. I. Fuks, A. Rubio, and N. T. Maitra, Phys. Rev.
Lett. {\bf 109}, 266404 (2012).

\bibitem{FERM13}
J. I. Fuks, P. Elliott, A. Rubio, and N. T. Maitra, J. Phys.
Chem. Lett. {\bf 4}, 735 (2013).

\bibitem{RG12}
J. D. Ramsden and R.W. Godby, Phys. Rev. Lett. {\bf 109}, 036402 (2012).


\bibitem{WB06}
F. Wilken and D. Bauer, Phys. Rev. Lett. {\bf 97}, 203001 (2006). 

\bibitem{WB07}
F. Wilken and D. Bauer, Phys. Rev. A. {\bf 76}, 023409 (2009). 

\bibitem{RHCM09}
A. K. Rajam, P. Hessler, C. Gaun, and N. T. Maitra, J. Mol. Struct. Theochem, {\bf 914}, 30 (2009). 


\bibitem{Henkel09}
N. Henkel, M. Keim, H. J. L\"udde, and T. Kirchner, Phys. Rev. A {\bf 80}, 032704 (2009). 

\bibitem{LFSEM14}
K. Luo, J. I. Fuks, E. Sandoval, P. Elliott, and N. T. Maitra, 
J. Chem. Phys. {\bf 140}, 18A515 (2014).

\bibitem{HK84}
M. F. Herman and E. A. Kluk, Chem. Phys. {\bf 91}, 27 (1984).

\bibitem{KHD86}
E. Kluk, M. F. Herman, and H. L. Davis, J. Chem. Phys. {\bf 84}, 326 (1986).

\bibitem{GX98}
F. Grossmann and A. L. Xavier, Phys. Lett. A {\bf 243} 243 (1998). 

\bibitem{EGCM11}
P. Elliott, S. Goldson, C. Canahui, and N.T. Maitra, Chem. Phys. {\bf 391}, 110 (2011).

\bibitem{FLSM15} J. I. Fuks, K. Luo, E. D. Sandoval, N. T. Maitra,
%{\it Time-resolved spectroscopy in time-dependent density functional theory: An exact condition}, 
 Phys. Rev. Lett. {\bf 114}, 183002 (2015). 

\bibitem{YNIB06}
K. Yabana, T. Nakatsukasa, J.-I. Iwata, G. F. Bertsch, Phys. Stat. Sol. (b) {\bf 243}, 1121 (2006). 


\bibitem{MZCB04}
N.T. Maitra, F. Zhang, R.J. Cave, and K. Burke, J. Chem. Phys. {\bf 120}, 5932 (2004). 

\bibitem{EM12}
P. Elliott and N. T. Maitra, 
%{\it Propagation of Initially Excited States in Time-Dependent Density-Functional Theory}, 
Phys. Rev. A. {\bf 85}, 052510 (2012).

\bibitem{MB01}
N.T. Maitra and K. Burke,  Phys. Rev. A {\bf 63}, 042501 (2001); {\bf 64} 039901 (E).



\end{thebibliography}
